\documentclass[12pt,a4paper,fleqn,twoside]{phart}
\usepackage{amssymb,amsmath,color}
\usepackage{graphicx,overcite,pxfonts}

\definecolor{gry}{rgb}{0.3,0.3,0.3}
\renewcommand\leq\leqslant
\renewcommand\geq\geqslant

\begin{document}

\thispagestyle{empty}

\noindent\hspace{0.08\linewidth}
\begin{minipage}[l]{0.92 \linewidth}
  
  \vspace{3cm}
  
  \noindent\LARGE\textsf{\textbf{\textcolor{gry}{Discovery and analysis of\\
      biochemical subnetwork hierarchies}}}

  \vspace{1cm}
  
  \noindent
  \large\sffamily\textbf{Petter Holme}\newline Department of Physics, Ume{\aa}
    University\newline 901$\,$87 Ume{\aa}, Sweden\vspace{0.35cm}
    
    \textbf{Mikael Huss}\newline SANS, NADA, Royal Institute of
  Technology\newline100$\,$44 Stockholm, Sweden

  \vspace{1cm}
  
  \normalsize\noindent\textbf{Abstract}
  
  \rmfamily\noindent
  The representation of a biochemical network as a graph is the
  coarsest level of description in cellular biochemistry. By studying
  the network structure one can draw conclusions on the large scale
  organisation of the biochemical processes. We describe methods how
  one can extract hierarchies of subnetworks, how these can be
  interpreted and further deconstructed to find autonomous
  subnetworks. The large-scale organisation we find is characterised
  by a tightly connected core surrounded by increasingly loosely
  connected substrates.
\end{minipage}

\pagestyle{myheadings}
\markboth{Holme \& Huss\hspace{3.5mm}
  \textit{Biochemical subnetwork
  hierarchies}}{\textit{Biochemical subnetwork
  hierarchies}\hspace{3.5mm} Holme \& Huss}

\section{Introduction}

At the coarsest level of description, cellular biochemistry can be
represented as a network of vertices (substrates) linked by chemical
reactions. For both conceptual and analytical purposes, the vastness
and complexity of these biochemical networks calls for a division into
smaller subunits. This is nothing new---traditionally biochemists have
talked about functional subnetworks, the citric acid cycle being one
example, comprised of biochemical pathways. As modern day genomics
gives an increasingly comprehensive picture of the biochemical network
one would like to complement the traditional way of mapping out
subnetworks by objective graph theoretical methods. By such methods
we can address not only the question what relevant subnetworks there
are, but also the hierarchical organisation of subnetworks (can
subnetworks be said to consist of smaller subnetworks, and so on), and
also more fundamental questions about in what context the subnetwork
concept is relevant and when the biochemical circuitry is to be
considered as a functional whole.

The graph-theoretical signature for a subnetwork is that it is
internally densely connected but has relatively few links to the rest
of the graph. Other methods for detecting
subnetworks~\cite{schus:dec,patra,john:hcs,bara:modhie} have been
based on local properties such as the number of reactions a substrate
takes part in, or the similarity of the neighbourhood. Since non-local
features can heavily affect network dynamics~\cite{holme:traffic},
one would prefer methods that take these into account. Here, we
discuss global algorithms for subnetwork detection, in particular
methods based on the betweenness centrality measure.

\section{Preliminaries}

\subsection{Biochemical networks as bipartite graphs}

A \textit{bipartite} graph\footnote{Or, to be precise, a
    \textit{two-mode representation} of a bipartite graph. The formal
    definition of bipartiteness is just that a graph contains no odd
    circuits.} contains of two types of vertices and links
that only go between vertices of different type.
We represent the biochemical networks as directed bipartite graphs
$G=(S,R,L)$ where $S$ is a set of vertices representing substrates, $R$
is a set of vertices representing chemical reactions, and $L$ is the
set of directed links---ordered pairs of one vertex in $S$ and one vertex
in $R$. The links are such that if the substrates $s_1,\cdots,s_n$ are
involved in a reaction $r\in R$ with products
${s'}_1,\cdots,{s'}_{n'}\in S$, then we have
$(s_1,r),\cdots,(s_n,r)\in L$ and $(r,{s'}_1),\cdots, (r,{s'}_{n'})\in
L$. The number of links leading to a vertex is called
\textit{in-degree} and denoted $k_\mathrm{in}$.

\subsection{Betweenness centrality}

Roughly speaking, the betweenness centrality~\cite{antonis} $C_B$ of a
vertex $v$ in an undirected graph is the number of shortest paths
between pairs of vertices that passes $v$. For the
purposes of this work we are interested in reaction vertices that are
central for paths between metabolites or other molecules; thus we
restrict our definition of betweenness to the reaction vertices
only. The precise definition then becomes:
\begin{equation}
  \label{eq:CB}
  C_B(r)=\sum_{s\in S}\sum_{s'\in S\setminus\{s\}}\frac{\sigma_{ss'}(r)}
  {\sigma_{ss'}}~,
\end{equation}
where $\sigma_{ss'}(r)$ is the number of shortest paths between $s$ and
$s'$ that passes through $r$, and $\sigma_{ss'}$ is the total number of
shortest paths between $s$ and $s'$. Since all substrates needs to be
present for a reaction to occur it is meaningful to rescale the
betweenness by the in-degree:
\begin{equation}
  \label{eq:CBeff}
  c_B(r)=C_B(r)/k_\mathrm{in}(r)~.
\end{equation}
We call $c_B$ the \textit{effective betweenness} of $v$.

\subsection{Girvan and Newman's algorithm}

The algorithm for tracing subnetworks we use is due to Girvan and
Newman (GN)~\cite{gir:alg}, but in a form adapted to bipartite
representations of biochemical networks as presented in
Ref.~\citen{hhj:sub}. The idea of the algorithm is based on the fact
that vertices that lie between densely connected areas have high
betweenness, and vice versa. Thus by successively removing reaction
vertices with high degree one will see the network disintegrate into
subnetworks of decreasing size. Furthermore, the smaller subnetworks
remaining after many iterations will be perfectly contained
subnetworks earlier in the execution of the algorithm, thus the method
produces a full hierarchy of subnetworks.

The precise definition of the algorithm is to repeat the following
steps until no reaction vertices remain:
\begin{enumerate}
\item Calculate the effective betweenness $c_B(r)$ for all reaction
  vertices.
\item\label{step2} Remove the reaction vertex with highest effective
  betweenness and all its in- and out-going links.
\item\label{step3} Save information about the current state of the
  network.
\end{enumerate}
If many reaction vertices have the same $c_B$ in step~\ref{step2},
we remove all of them at once. A C-implementation of this algorithm
along with test data sets can be found at
\texttt{www.tp.umu.se/forskning/networks/meta/}.

\section{A case-study: \textit{T.\ pallidum}}

\begin{figure}\label{fig:tree}
  \centering{\resizebox*{\textwidth}{!}{\includegraphics{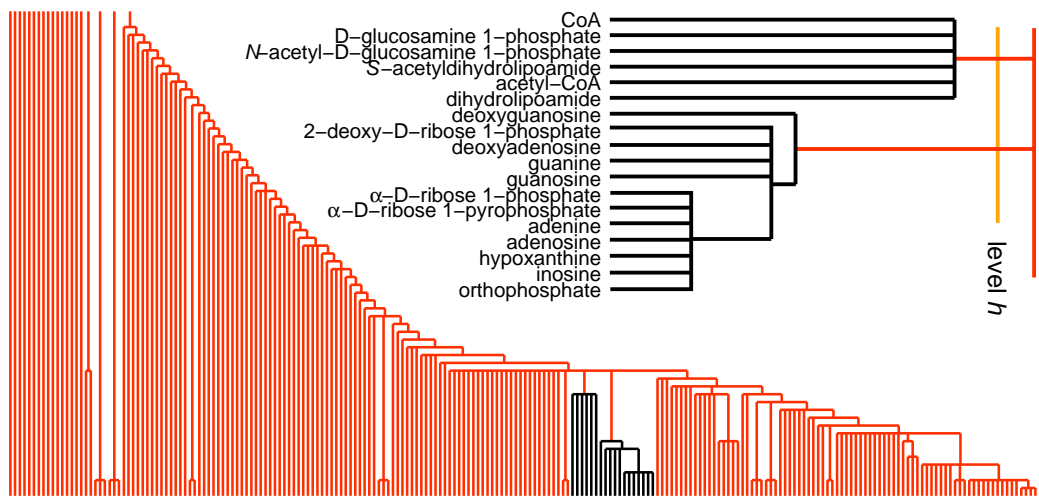}}}
\caption{The hierarchical clustering tree for the metabolic network of
  \textit{T.\ pallidum}. The inset shows substrate names for a blow-up
  of the tree (indicated by black).}
\end{figure}

To illustrate the output of the algorithm, and how it can be
post-processed, we choose the metabolic network of \textit{T.\
  pallidum}---the pathological agent of syphilis---as obtained from
the WIT database~\cite{wit}\footnote{This is the same data as used in
  Refs.~\citen{jeong:meta,bara:modhie}, and thus slightly outdated, but
  it should work well for illustrating the method.}.

\subsection{The large scale shape of the hierarchy trees}

The subnetwork hierarchy of \textit{T.\ pallidum}'s metabolic network
is presented as a tree (a so-called \textit{dendrogram}) in
Fig.~\ref{fig:tree}. The end-points at the base of the
dendrogram represent the substrate vertices of the metabolic
network. The vertical dimension represents the hierarchical level---if
a horizontal line is drawn across the dendrogram, the vertices
connected below the line belongs to the same cluster (connected
subgraph) at that particular level of the hierarchical
organisation. The further down the tree two vertices are connected,
the more tightly connected are they in the biochemical network. If one
substrate is to be converted to another that is separated from the
first one high up in the in the dendrogram, then a long chain of
reactions is needed. If, on the other hand, the two vertices are
connected near the bottom of the dendrogram, then they probably are
both present in one or more reactions.

The most striking feature of Fig.~\ref{fig:tree}, and indeed of any of the
43 organisms of Ref.~\citen{jeong:meta}, is that the network has one
dominating cluster at most levels of the hierarchy. As the algorithm
proceeds (one goes from top to bottom of the dendrogram) a few
vertices at a time peel off from the largest connected cluster. The
emerging picture is that the large scale structure of metabolic
network has a tightly connected core and increasingly loosely
connected outer `shells.' A few rather well-defined sub-networks are
identified however, for example the subnetworks of Fig.~\ref{fig:tree}
containing reactions associated with purine metabolism and
pyruvate/acetyl-CoA conversion.

\subsection{Criteria for identifying subnetworks}

We can identify subnetworks by looking at the hierarchy tree, if a
subnetwork is isolated at some level (like the
\textit{N}-acetyl-D-glu\-cos\-amine 1-phos\-phate, D-glu\-cos\-amine
1-phos\-phate, dihydrolipoamide,
\textit{S}-ace\-tyl\-di\-hydro\-lipo\-amide, CoA, and acetyl-CoA
network of Fig.~\ref{fig:tree} at level $h$) then it is comparatively
well connected within itself relative to its surrounding. If the cluster is
isolated close to the top of the dendrogram, then it is not very
entangled in the wirings of metabolic pathways, and likely to be a
reasonably autonomously functioning module. Can we establish objective
criteria for subnetworks to be regarded as meaningful modules?
For example Ref.~\citen{bara:modhie} detects modules in an indirect
way using a very weak criterion, roughly speaking, that substrates
are likely to belong to same module if they appear in reactions
involving the same set of other substrates. To identify groups in
social networks Radicci \textit{et al.}\cite{castel:comm}\ suggested two
criteria that, adapted to biochemical networks becomes as
follows: If, during the iterations of the GN algorithm, an
isolated vertex set $S'\subset S$ fulfils the following criterion it
is said to be a \textit{weak community}:
\begin{equation}\label{eq:weak}
  \sum_{s\in S'}K_\mathrm{in}(s) >  \sum_{s\in S'}K_\mathrm{out}(s)~,
\end{equation}
and a \textit{strong community} if:
\begin{equation}\label{eq:strong}
  K_\mathrm{in}(s) >  K_\mathrm{out}(s) \mbox{~for all~} s\in S'~,
\end{equation}
where $K_\mathrm{in}(s)$ is the number of $s\in S$ that are
products of a reaction involving a substrate $s\in S$, and
$K_\mathrm{out}(s)$ is the number of $s\in S\setminus S'$ that are
products of a reaction involving a substrate $s\in S$. Loosely
speaking Eq.~\ref{eq:weak} means that there are, on average, more
feedback pathways back into $S'$ than pathways leading out to the
rest of the network. If the strong condition (Eq.~\ref{eq:strong})
holds, then products of all reactions involving substrates $s\in S'$
are more likely to belong to $S'$ than not. It turns out that
Eq.~\ref{eq:strong} is not fulfilled for almost any cluster at any
but the lowest level of the hierarchy (closest to the bottom of the
dendrogram).  Eq.~\ref{eq:weak} is on the other hand fulfilled for
the largest cluster throughout all iterations of the algorithm. (This
picture persists for all 43 WIT organisms studied in
Ref.~\citen{jeong:meta}.) That the subnetworks of cellular biochemistry
almost completely lacks the community structure of social network, or
component structure of electronic devices, does not necessarily mean
that it is futile to talk of biochemical modules. For a subnetwork to
have some degree of autonomy it has to have some self-regulatory
function, and thus a feedback loop. To implement this idea, consider
the subnetworks with substrate vertex set $S'$ that fulfils:
\begin{equation}
  L(S') \leq  \Lambda|S'|~,\label{eq:l}
\end{equation}
where $L(S')$ is the number of vertices in $S'$ that lies on an
elementary cycle (a closed non-self-intersecting path) of only vertices
in $S'$ and length larger than three, $|S'|$ is the number of
vertices in $S'$, and the parameter $\Lambda\in[0,1]$ is the required
fraction of feedback loop vertices. We test the three cases where
$\Lambda$ equals $0$, $1/2$ and $1$, corresponding to the subnetwork
having at least one feedback loop, more than half of the substrates,
or every substrate participating in a feedback loop,
respectively. The largest cluster close to the top of the dendrogram
quite naturally fulfils Eqs.~\ref{eq:l} when $\Lambda$ small (in our
case $0$ or $1/2$), therefore we detect subnetworks starting from the
bottom of the dendrogram and go upwards.  With each one of these
criteria we find non-trivial subnetworks. Of the subnetworks of
Fig.~\ref{fig:tree} the hardest requirement, $\Lambda = 1$ detects two
relevant subnetworks---the one containing CoA and the innermost one
containing orthophosphate: $\alpha$-D-ribose 1-phosphate,
$\alpha$-D-ribose 1-pyrophosphate, adenine, adenosine, hypoxanthine,
inosine, and orthophosphate. The extended ortophosphate-subnetwork
still connected at level $h$ (also containing e.g.\ guanine) is
regarded as a valid subnetwork with $\Lambda = 1/2$, but not with
$\Lambda = 1$. To assign an appropriate $\Lambda$ requires a careful
look at the problem in question, but as a rule of thumb $\Lambda$
close to one seems sensible for most applications.

\section{Conclusions}

Finding subnetworks of cellular biochemistry is an important task for
modern bioinformatics, for both conceptual and analytical
purposes. There are two general ways to proceed, either one searches
for small building blocks (cf.\ Ref.~\citen{alon}) or one tries to
deconstruct the whole network. Our approach falls into the second
category. By adapting an algorithm~\cite{gir:alg} for subnetwork
detection to biochemical networks we construct hierarchy trees,
dendrograms,  representing the whole hierarchical organisation of
subnetworks of biochemical pathways. We find that biochemical networks
cannot be divided into subnetworks as easily as e.g.\ acquaintance
networks, and electronic circuits~\cite{hhj:sub}. Against this
backdrop it is not surprising that some recent criteria
(Eqs.~\ref{eq:weak} and \ref{eq:strong}) for extracting meaningful
social subnetworks fail to give non-trivial results. In remedy we
propose conditions based on the presence of feedback loops within a
subnetwork. The above methods are illustrated by an application to the
metabolic network of \textit{T.\ pallidum}, we have also tested them on
the metabolic and whole-cellular networks (containing e.g.\
transmembrane transport and signal transduction) of 42 other organisms
of the WIT database~\cite{wit}, and obtain sensible output.

\subsection*{Acknowledgements}
Thanks are due to Claudio Castellano, Hawoong Jeong and Petter
Minn{\-}hagen. P.H.\ was partially supported by Swedish Research
Council through contract no.\ 2002-4135.

\end{document}